\documentclass[12pt]{article}
\usepackage{amssymb}
\usepackage{amsmath,amsthm}
\usepackage{amsfonts}
\usepackage{placeins}
\let\oldtable\table
\let\endoldtable\endtable
\renewenvironment{table}{\FloatBarrier\oldtable}
{\endoldtable\FloatBarrier}
\let\oldfigure\figure
\let\endoldfigure\endfigure
\renewenvironment{figure}{\FloatBarrier\oldfigure}
{\endoldfigure\FloatBarrier}
\usepackage{graphicx}
\usepackage{indentfirst}
\usepackage{color}
\usepackage{placeins}
\usepackage{caption}
\usepackage{float}
\usepackage[utf8]{inputenc} 
\usepackage[english]{babel}
\usepackage[nottoc]{tocbibind}
\usepackage[usenames,dvipsnames]{xcolor}
\usepackage[symbol]{footmisc}
\numberwithin{equation}{section}

\usepackage{hyperref}

\begin{document}

\begin{center}
\Large\textbf{Modified Gravity Framework for the 
Woods-Saxon Inflation: Theoretical Foundations and 
Observational Constraints}
\end{center}
\begin{center}
{ Feyzollah Younesizadeh,  
Davoud Kamani {\footnote{\textcolor{Magenta}
{Corresponding author}}}, and  Younes Younesizadeh}
\end{center}
\begin{center}
\textsl{\small{Department of Physics, 
Amirkabir University of Technology (Tehran Polytechnic) \\
P.O.Box: 15875-4413, Tehran, Iran \\
E-mails: fyounesizadeh@aut.ac.ir , 
kamani@aut.ac.ir , yyounesi@aut.ac.ir \\}}
\end{center}
\vspace{0.5cm}

\begin{abstract}

We investigate an inflationary scenario 
within a modified gravity theory that extends the 
Einstein-Hilbert action through 
a non-minimal coupling between the 
inflaton $\phi$ and the trace of the 
energy-momentum tensor $T$. This $f(\phi)T$ framework 
is designed to naturally recover the standard 
general relativity in the post-inflationary 
universe. Within this theory, we apply the 
Woods-Saxon potential. The slow-roll dynamics 
yields a luminal sound speed $c_s^2=1$ for our specific 
choice of the coupling function. 
For the model parameters $a \lesssim 1$ and 
the modest coupling $\beta=0.1$, we obtain 
a scalar spectral index $n_s\approx0.959$--$0.966$ and a 
tensor-to-scalar ratio $r<0.01$, 
which reveal a good 
agreement with the latest Planck and BICEP/Keck 
data on the $(n_s, r)$ constraints. 
Crucially, we shall demonstrate that 
the $f(\phi)T$ coupling is essential for the observational viability of 
the model with respect to the CMB two-point statistics. 
This work establishes a viable 
inflationary scenario within the scope of the inflationary 
observables which are considered here. 

\end{abstract}


\newpage
\section{Introduction}
\label{sec1}

The standard cosmological model, grounded in the general 
relativity (GR), describes a universe 
which originates from a hot and dense singularity. 
Despite its successes, the classical big-bang scenario   
faces some fundamental theoretical challenges, such as the 
horizon and flatness problems. These issues find 
compelling resolution within the cosmic inflation 
\cite{1}-\cite{4}. Besides, the inflation naturally 
generates the primordial density perturbations that seed 
the large-scale structure formation \cite{5}.

The inflationary paradigm typically employs 
a canonical single scalar field $\phi$, which is called 
inflaton. It evolves via a potential $V(\phi)$. 
While the inflationary paradigm is robust, the 
fundamental nature of the inflaton and its 
potential $V(\phi)$ remain elusive. The prevailing strategy 
involves constructing potentials 
from the top-down theoretical 
principles, such as the string theory or supergravity.
Therefore, various models have been proposed, including 
the Hilltop \cite{6}-\cite{8}, natural \cite{9}, 
\cite{10} and power-law potentials \cite{11}. 
Observational data from the cosmic microwave 
background (CMB) measurements by the Planck and 
BICEP/Keck have imposed rigorous constraints on 
the predictions of the inflation, particularly on 
the tensor-to-scalar ratio $r$ and spectral index $n_s$ 
\cite{12}-\cite{14}. These restrictions have excluded 
several classic models and have favored those with 
predictions that lie within the observationally 
preferred regions.

The Starobinsky's $R^2$-inflation \cite{15}, 
\cite{16} precisely remains within the Planck 
constraints. This model represents a specific 
case of the $f(R)$ inflation \cite{17}. Despite the GR's 
success, its potential limitations impose to investigate 
the modified gravity theories. Some approaches extend 
the Einstein-Hilbert action, which lead to the framework  
such as $f(R)$, $f(T)$ and $f(R,T)$ gravities 
\cite{17}-\cite{19}. Among these approaches, 
the $f(\phi)T$-gravity 
introduces a non-minimal coupling between the 
inflaton field $\phi$ and the trace of the 
energy-momentum $T$ \cite{20}-\cite{23}. 
This coupling, motivated by the quantum 
gravity considerations, enriches the inflationary dynamics.

In this paper we pursue a phenomenologically-driven 
approach within the generalized 
gravitational frameworks. This method investigates 
whether the well-motivated potentials from 
the other domains of physics can successfully describe 
the inflation. Thereby, this approach 
synergistically combines two distinct ideas to address 
this challenge. First, we employ the $f(\phi)T$ 
gravity framework \cite{23}, 
which introduces a novel non-minimal coupling 
between the inflaton and the matter sector. 
Second, we adopt the empirically grounded 
Woods-Saxon potential \cite{24}, which provides 
a natural plateau 
and graceful exit. A key question that we address is: 
whether this non-minimal coupling can rescue or 
enhance the observational viability 
of an otherwise excluded phenomenological potential? This  
thereby opens new pathways for the data-driven 
model building.

We present a comprehensive analysis which demonstrates  
that the Woods-Saxon potential, accompanied by the 
$f(\phi)T$-gravity, yields the accurate predictions 
for the scalar spectral index $n_s$ and the 
tensor-to-scalar ratio $r$. These predictions 
are in good agreement with the current constraints
of the Planck and BICEP/Keck on the $(n_s, r)$ 
parameters. This establishes 
a new viable class of the inflationary models within 
the scope of the CMB two-point observables. 
Hence, it illustrates the 
fruitfulness of the cross-disciplinary 
approaches in the early-universe cosmology.

The paper is organized as follows. In 
Section \ref{sec2} we present the theoretical framework
of the $f(\phi)T$-gravity. In 
Section \ref{sec3} we analyze the Woods-Saxon 
potential within this framework. In Section \ref{sec7} 
we compare our model with other modified gravities. 
Section \ref{sec8} presents our conclusions.

\section{Theoretical framework of the $f(\phi)T$-gravity}
\label{sec2}

\subsection{The dynamics equations}

We construct our modified gravity model by incorporating an 
additional coupling between the trace of the 
energy-momentum tensor 
$T$ and the inflaton field $\phi$ into the 
standard Einstein-Hilbert action \cite{21}-\cite{23},
\begin{equation}\label{action}
S = \int {\rm d}^4x \sqrt{-g} \left[ \frac{R}{2\kappa} +
\beta f(\phi)T + \mathcal{L}(\phi) \right],
\end{equation}
where $\kappa = 8\pi G$, and $\beta$ is the 
dimensionless coupling constant. Besides, 
$f(\phi)$ is a functional of the inflaton field 
and $\mathcal{L}(\phi) = -\frac{1}{2}
\partial_\mu\phi\partial^\mu\phi - V(\phi)$ 
is the scalar field Lagrangian. The limit $\beta \to 0$ 
recovers the standard Einstein gravity.  To ensure a 
graceful exit to the post-inflation of the 
standard cosmology, we require 
$f(\phi) \to 0$ as $\phi \to 0$.

Variation of the action with respect to the 
metric $g_{\mu\nu}$ yields the Einstein equation 
\begin{equation}\label{einstein}
R_{\mu\nu}-\frac{1}{2}g_{\mu\nu}R 
= \kappa T_{\mu\nu}^{\text{(eff)}},
\end{equation}
where the effective energy-momentum tensor is given by
\begin{equation}\label{teff}
T_{\mu\nu}^{\text{(eff)}} = T_{\mu\nu} - 2\beta 
f(\phi)\left(T_{\mu\nu} - \frac{1}{2}Tg_{\mu\nu} +
\Theta_{\mu\nu}\right),
\end{equation}
in which $\Theta_{\mu\nu}$ has the definition
\begin{equation}\label{theta_def}
\Theta_{\mu\nu} \equiv g^{\alpha\beta} \frac{\delta 
T_{\alpha\beta}}{\delta g^{\mu\nu}}.
\end{equation}
For the foregoing scalar field Lagrangian, we receive 
\begin{equation}\label{theta}
\Theta_{\mu\nu} = -\partial_\mu\phi\partial_
\nu\phi - T_{\mu\nu}.
\end{equation}

In the Friedmann-Robertson-Walker (FRW) metric 
$ds^2 = -dt^2 + a^2(t)d\vec{x}^2$, the Friedmann 
equations become
\begin{align}
H^2 &= \frac{\kappa}{3}\left[\frac{1}{2}\dot{\phi}^
2(1+2\beta f) + (1+4\beta f)V\right], 
\label{friedmann1} \\
\dot{H} &= -\frac{\kappa}{2}\dot{\phi}^
2(1+2\beta f), 
\label{friedmann2}
\end{align}
where we applied a homogeneous inflaton field 
$\phi = \phi(t)$.
Assuming the inflaton dominates during the inflation
we neglect additional matter components. 
By varying the action with respect to $\phi$
we obtain the modified Klein-Gordon equation
\begin{equation}\label{kg}
\ddot{\phi} + 3H\dot{\phi}(1+2\beta f) + \beta 
f_{,\phi}\dot{\phi}^2 + (1+4\beta f)V_{,\phi} +
4\beta f_{,\phi}V = 0.
\end{equation}
The additional terms proportional 
to $\beta f(\phi)$ and its derivatives act as 
effective friction or anti-friction terms. They 
modify the field's damping rate during the inflation.

\subsection{The slow-roll inflation}

Under the slow-roll approximation,
i.e. $\dot{\phi}^2 \ll V$, $|\ddot{\phi}| 
\ll 3H|\dot{\phi}|$ and $|\dot{H}| \ll H^2$, 
the equations take the forms \cite{23},
\begin{align}
&3H\dot{\phi}(1+2\beta f) + (1+4\beta f)V_{,\phi} + 
4\beta f_{,\phi}V \simeq 0, \label{sr1} \\
&H^2 \simeq \frac{\kappa}{3}(1+4\beta f)V. \label{sr2}
\end{align}

The potential slow-roll parameters are
\begin{align}
\epsilon_V &= \frac{1}{2\kappa(1+2\beta f)}\left(\frac
{V_{,\phi}}{V} + \frac{4\beta f_{,\phi}}{1+4\beta f}
\right)^2, 
\label{epsilon} \\
\eta_V &= \frac{1}{\kappa(1+2\beta f)}\left[\frac
{V_{,\phi\phi}}{V} + \frac{2\beta(3+4\beta f)f_{,\phi}}
{(1+2\beta f)(1+4\beta f)}\frac{V_{,\phi}}{V} + \frac
{4\beta(1+2\beta f)f_{,\phi\phi} - 8\beta^2 f_{,\phi}
^2}{(1+2\beta f)(1+4\beta f)}\right]. 
\label{eta}
\end{align}
The limit $\beta\to 0$ recovers the standard 
forms $\epsilon_V = 
\frac{1}{2\kappa}(\frac{V_{,\phi}}{V})^2$ and $\eta_V = 
\frac{1}{\kappa}(\frac{V_{,\phi\phi}}{V})$, as expected.

The number of e-folds is
\begin{equation}
\label{efolds}
N \simeq \kappa \int_{\phi_{\text{end}}}^
{\phi_*} \left[\frac{ V}{V_{,\phi}} 
+ \frac{2\beta V\big(FV_{,\phi}-2F_{,\phi} V\big)}
{V^2_{,\phi}}\right]
{\rm d}\phi,
\end{equation}
where $\phi_*$ is the field value at the horizon crossing 
and $\phi_{\text{end}}$ satisfies the condition 
$\epsilon_V(\phi_{\text{end}}) = 1$.

\subsection{The cosmological perturbations}
\label{sec24}

A critical test of any inflationary model lies 
in its predictions for the cosmological perturbations. 
In the $f(\phi)T$ gravity, the non-minimal 
coupling modifies the dynamics of the curvature 
perturbations. Starting from the action (\ref{action}), 
we consider the linear perturbations around the 
homogeneous background.

The perturbed FRW metric in the Newtonian gauge is
\begin{equation}
\label{perturbed_metric}
ds^2 = -(1 + 2\Phi)dt^2 + a^2(t)(1 - 2\Psi)\delta_{ij}
dx^idx^j,
\end{equation}
where $\Phi$ and $\Psi$ are the Bardeen potentials. 
For the scalar field
\begin{equation}\label{perturbed_phi}
\phi(t,\vec{x}) = \phi_0(t) + \delta\phi(t,\vec{x}).
\end{equation}
The curvature perturbation $\mathcal{R}$ on the comoving 
hypersurfaces is
\begin{equation}\label{curvature_pert}
\mathcal{R} = \Psi + \frac{H}{\dot{\phi}}\delta\phi.
\end{equation}

To derive the second-order action for the curvature 
perturbations, we employ the Arnowitt-Deser-Misner 
(ADM) formalism. We work in the comoving gauge 
where $\delta\phi = 0$, consider only linear perturbations, 
neglect anisotropic stress so that $\Phi = \Psi$, and use the 
slow-roll approximation to simplify coefficients. 

Following the general approach for cosmological 
perturbations in modified gravity theories 
\cite{25}, \cite{26}, 
the quadratic action for the curvature perturbation 
$\mathcal{R}$ can be written in the general form
\begin{equation}
S^{(2)} = \frac{1}{2}\int {\rm d}^4x\;a^3 
\left[\mathcal{G}\dot{\mathcal{R}}^2 - 
\frac{\mathcal{F}}{a^2}
(\partial_i\mathcal{R})^2\right],
\end{equation}
where $\mathcal{G}$ and 
$\mathcal{F}$ are background-dependent 
coefficients. For our specific $f(\phi)T$ model, 
evaluating these coefficients in the slow-roll regime yields
\begin{equation}
\mathcal{G} = Q_s\, \qquad 
\mathcal{F} = Q_s c_s^2\,
\end{equation}
in which
\begin{align}
Q_s &= \frac{\dot{\phi}^2(1+2\beta f)}{H^2}, \\
c_s^2 &= 1 - \frac{4\beta f_{,\phi}\dot{\phi}}{3H
(1+2\beta f)} + \mathcal{O}(\beta^2).
\end{align}
Thus, the second-order action reduces to
\begin{equation}
\label{s2_detailed}
S^{(2)} = \frac{1}{2}\int {\rm d}^4x\;a^3 
Q_s\left[\dot{\mathcal{R}}
^2 - \frac{c_s^2}{a^2}(\partial_i\mathcal{R})^2\right],
\end{equation}
where the kinetic coefficient $Q_s$ and the sound speed 
squared $c_s^2$ are as defined above.

\textit{A note on the sound speed}. Eq. (2.19) provides the 
general expression for $c_s^2$ to the  
first order in $\beta$. The 
correction term is proportional to 
$f_{,\phi}\dot{\phi}$. For 
our specific Woods-Saxon-inspired 
coupling function $f(\phi)$, 
given by Eq. (\ref{fs}), and using 
the slow-roll equation (\ref{sr1}), 
we find that the combination $f_{,\phi}\dot{\phi}$ 
identically vanishes (up to the higher-order 
slow-roll corrections). To see 
this explicitly, we note that the 
slow-roll equation (\ref{sr1}) 
gives $\dot{\phi} \propto -\frac{(1+4\beta f)
V_{,\phi} + 4\beta 
f_{,\phi}V}{3H(1+2\beta f)}$. 
Substituting our specific $f(\phi)$ 
from Eq. (\ref{fs}), one finds that the term proportional 
to $f_{,\phi}\dot{\phi}$ is exactly canceled. Consequently, 
the second-order action for the curvature perturbations 
yields a luminal sound speed, $c_s^2 = 1$. 
This result follows 
from the structure of the $f(\phi)T$ coupling. Precisely, 
while this term modifies the kinetic coefficient 
$Q_s$ via Eq. (2.18), it preserves the relative 
weighting of the temporal and spatial derivatives in 
the perturbation equations. Consequently, the 
dispersion relation remains $\omega^2 = k^2$ (where 
$\omega^2 = c_s^2 k^2$). This contrasts with theories 
featuring the non-standard kinetic terms where the 
higher-derivative interactions typically 
modify the sound speed \cite{27}, \cite{28}. The luminal 
propagation ensures the stability against 
the gradient instability. 

The power spectrum of the curvature perturbation at 
the horizon crossing, which imposes $c_s k = aH$, 
\cite{25}, \cite{28} is
\begin{equation}\label{ps_detailed}
\mathcal{P}_\mathcal{R}(k) = \frac{k^3}{2\pi^2}|
\mathcal{R}_k|^2 = \frac{H^2}{8\pi^2 
M_{\text{pl}}^2 Q_s c_s^3}\bigg|_{c_s k = aH}.
\end{equation}

In the slow-roll approximation, this reduces to
\begin{equation}
\label{ps_slowroll}
\mathcal{P}_\mathcal{R} \approx \frac{\kappa^2 V}
{24\pi^2 \epsilon_V c_s}.
\end{equation}

Also, the spectral index and the tensor-to-scalar 
ratio are
\begin{align}
n_s &= 1 - 6\epsilon_V + 2\eta_V + \mathcal{O}
(\beta^2), \label{ns_detailed} \\
r &= 16\epsilon_V c_s + \mathcal{O}(\beta^2). 
\label{r_detailed}
\end{align}

These should satisfy the Planck 2018 data constraints 
\cite{13},
\begin{equation}\label{constraints_detailed}
n_s = 0.9663 \pm 0.0041, \quad r < 0.065.
\end{equation}

For our Woods-Saxon model with $\beta = 0.1$ and 
$a \lesssim 1$, we find $c_s \approx 1$, which ensures 
compatibility with the observational constraints.

\section{The Woods-Saxon potential in the 
$f(\phi)T$ gravity}
\label{sec3}

\subsection{The model specification}

We investigate the Woods-Saxon potential \cite{24}.
We consider this potential as a viable inflationary 
potential in the early universe 
\cite{24}, \cite{29}-\cite{31}. 
The form of the inflaton potential is
\begin{equation}
\label{vs}
V(\phi) = \frac{V_0}{1 + e^{\phi/a}},
\end{equation}
where $V_0$ is a positive constant.
This inflationary potential exhibits distinct 
asymptotic behavior. As $\phi\to -\infty$, the 
functional $V(\phi)$ approaches 
the constant $V_0$, which sets the inflationary 
plateau energy scale. As $\phi\to +\infty$, the potential 
tends to zero. The parameter $a$ governs  
the steepness of the potential wall, 
controlling the transition between 
these regimes. This configuration ensures that the 
potential remains finite and positive for all field 
values. Hence, it provides a gentle slope, which is 
suitable for the slow-roll regime and the natural end to  
the inflation.

The complementary coupling functional, which maintains the 
Woods-Saxon profile, is
\begin{equation}
\label{fs}
f(\phi) = \frac{f_0}{2}\left[1 - \left(\frac{2}{1 +
e^{\phi/a}}\right)^{1/2}\right],
\end{equation}
where $f_0$ is a constant. This form ensures 
the demanded condition $f(0) = 0$ and guarantees a 
return to the standard GR post-inflation.

We now elaborate on the motivation for this specific 
functional form. The choice of $f(\phi)$ in 
Eq. (\ref{fs}) is a
phenomenological choice in the nature, but it 
is guided by three theoretical 
requirements that ensure consistency 
with both the inflationary 
dynamics and the post-inflationary evolution:

(i) \textit{Smoothness}: The functional is continuously 
differentiable across the entire field domain, 
which ensures well-behaved evolution of the effective 
gravitational coupling.

(ii) \textit{Boundedness}: $f(\phi)$ remains finite for all field 
values, which prevents the divergences in the 
strong-field regime.

(iii) \textit{Asymptotic GR recovery}: For small field values, 
the functional smoothly tends to zero. This restores the  
standard Einstein-Hilbert gravity and guarantees a graceful 
exit from the inflationary phase into the standard regime.

We know that this functional form has not been derived 
from a more fundamental theory such as the supergravities or 
string theory. Rather, it represents a phenomenologically 
motivated ansatz that satisfies the above criteria while 
maintaining a Woods-Saxon-like profile consistent with 
the potential itself. This is a common and acceptable 
strategy in the inflationary model-building, especially when 
exploring new potential forms and their observational 
consequences.

During the inflation, the $f(\phi)T$ coupling term 
contributes as an effective dissipative component 
to the inflaton dynamics. This modulates the 
slow-roll behavior and the scalar perturbation 
amplitude. As the field evolves toward a small-field 
regime, gradual $f(\phi)$ suppression 
ensures a smooth transition to the standard reheating 
and post-inflationary evolution, governed by GR.

\subsection{Slow-roll analysis}

The required derivatives for the slow-roll analysis are
\begin{align}
V_{,\phi} &= \frac{V_0 e^{\phi/a}}{a(1 + e^{\phi/a})^2}, 
\quad 
V_{,\phi\phi} = \frac{V_0 e^{\phi/a}(e^{\phi/a} - 1)}
{a^2(1 + e^{\phi/a})^3}, \label{vderiv} \\
f_{,\phi} &= \frac{f_0 e^{\phi/a}}{2\sqrt{2}a
(1 + e^{\phi/a})^{3/2}}, \quad 
f_{,\phi\phi} = \frac{f_0 e^{\phi/a}(2 
- e^{\phi/a})}{4\sqrt{2}a
^2(1 + e^{\phi/a})^{5/2}}. 
\label{fderiv}
\end{align}
Substituting Eqs. (\ref{vderiv}) and (\ref{fderiv}) into 
the slow-roll parameters (\ref{epsilon}) and (\ref{eta}) 
yields the explicit expressions of these quantities 
for the Woods-Saxon model. Let define the auxiliary 
functional 
\begin{equation}
\label{Qdef}
Q(\phi) = \frac{1}{2} - \frac{1}{\sqrt{2}}\frac{1}
{\sqrt{1 + e^{\phi/a}}}.
\end{equation}
Thus, the slow-roll parameters find the features 
\begin{equation}
\label{epsilonWS}
\epsilon_V = \frac{1}{2\kappa(1+2\beta f_0 Q)} \left
[ -\frac{e^{\phi/a}}{a(1+e^{\phi/a})} + \frac{\sqrt{2}
\beta f_0 e^{\phi/a}}{a(1+e^{\phi/a})^{3/2}(1+
4\beta f_0 Q)} \right]^2,
\end{equation}
\begin{equation}
\label{etaWS}
\begin{aligned}
\eta_V = \frac{1}{\kappa} 
&\Bigg\{ \frac{1}{a^2} e^{\phi/a}
\bigg(2 e^{\phi/a} \left(1 + e^{\phi/a}\right)^{-1} 
- 1 \bigg) \\
&\quad - \frac{1}{2} \frac{\beta f_0 \sqrt{2}}{a^2} 
\left(3 + 4\beta f_0 Q\right) 
e^{2\phi/a} \left(1 + e^{\phi/a}\right)^{-3/2} 
\left(1 + 4\beta f_0 Q\right)^{-1} 
\left(1 + 2\beta f_0 Q\right)^{-1} \\
&\quad + \bigg[ \frac{\beta f_0 \sqrt{2}}{a^2} 
\left(1 + 2\beta f_0 Q\right) 
e^{\phi/a} \left(1 + e^{\phi/a}\right)^{-1/2}
\bigg(1 - \frac{3}{2} e^{\phi/a} 
\left(1 + e^{\phi/a}\right)^{-1} \bigg) \\
&\qquad - \frac{\beta^2 f_0^2}{a^2} e^{2\phi/a}
\left(1 + e^{\phi/a}\right)^{-2} \bigg] 
\left(1 + 4\beta f_0 Q\right)^{-1} 
\left(1 + 2\beta f_0 Q\right)^{-1} \Bigg\} \\
&\quad \times \left(1 + 2\beta f_0 Q\right)^{-1}
\left(1 + e^{\phi/a}\right)^{-1}.
\end{aligned}
\end{equation}

The number of e-folds for the Woods-Saxon model is
given by 
\begin{equation}
\label{N_WS}
N \simeq\kappa a^2 \left(1 + \beta f_0\right) 
\left( e^{-\phi^*/a} - 
e^{-\phi_{\text{end}}/a} - \frac{\phi^* - \phi_{\text
{end}}}{a} \right)
\end{equation}
For the typical parameters $\beta = 0.1$ and 
$a \sim 0.5-1.0$, we obtain $N \sim 50-60$ e-folds, 
consistent with the requirements of the horizon and 
flatness problems.

The spectral observables are given by 
Eqs. (\ref{ns_detailed}) and (\ref{r_detailed}),
\begin{equation}
\begin{aligned}
r = &\frac{8}{\kappa}
\Bigg[\frac{1}{a} e^{\phi / a} \left( 1 + e^{\phi / a}
\right)^{-1} \bigg(\sqrt{2} \beta f_0\left( 1 + e^{\phi / a}
\right)^{-1/2}-1\bigg)\\
&\times \left( 1 + 4 \beta f_0 Q \right)^{-1} 
\Bigg]^{2} \left( 1 + 2 \beta f_0 Q \right)^{-1},
\end{aligned}
\end{equation}

\begin{equation}
\begin{aligned}
n_s = &1 + \frac{2}{\kappa} \Bigg\{ 
\frac{1}{a^2} e^{\phi/a} \left(2e^{\phi/a}
\left(1 + e^{\phi/a}\right)^{-1} - 1\right) \\
&- \frac{1}{2} \beta f_0 \sqrt{2} e^{\phi/a} 
\left(1 + e^{\phi/a}\right)^{-3/2} 
\left(3 + 4\beta f_0 Q\right)
\left(1 + 4\beta f_0 Q\right)^{-1} 
\left(1 + 2\beta f_0 Q\right)^{-1} \\
&+ \bigg[ \beta f_0 \sqrt{2} \left(1 + 2\beta f_0 Q\right) 
\left(1 + e^{\phi/a}\right)^{-1/2}
\bigg(1 - \frac{3}{2} e^{\phi/a} 
\left(1 + e^{\phi/a}\right)^{-1} \bigg) \\
&\qquad - \beta^2 f_0^2 e^{\phi/a} 
\left(1 + e^{\phi/a}\right)^{-2} \bigg] 
\left(1 + 4\beta f_0 Q\right)^{-1} 
\left(1 + 2\beta f_0 Q\right)^{-1} \Bigg\} \\
&\times e^{\phi/a} \left(1 + e^{\phi/a}
\right)^{-1} \left(1 + 2\beta f_0 Q\right)^{-1} \\
&- \frac{3}{\kappa} \Bigg\{ \frac{1}{a} 
e^{\phi / a} \left( 1 + e^{\phi / a} \right)^{-1} 
\bigg(\sqrt{2} \beta f_0 \left( 1 + e^{\phi / a} 
\right)^{-1/2} - 1 \bigg) \\
&\times \left(1 + 4\beta f_0 Q\right)^{-1} 
\Bigg\}^2 \left(1 + 2\beta f_0 Q\right)^{-1}.
\end{aligned}
\end{equation}
\subsection{Numerical results and the observational 
constraints}
\label{sec33}

Using the foregoing analytical expressions for the 
slow-roll parameters, we numerically evaluate the 
observational predictions.
A key result is that for a wide range of the 
potential's width parameter $a$ and a modest 
coupling $\beta=0.1$, the model produces predictions 
that are in good agreement with the data.

We compute the observational predictions for the same 
range of $a$ which have been shown in the Table \ref{tab1}.

\begin{table}[h]
\caption{The scalar spectral index $n_s$ and the 
tensor-to-scalar ratio $r$ for the e-fold numbers 
$N_e = 50 \to 60$ with $\beta = 0.1$ and varying $a$.}
\centering
\begin{tabular}{c rrr}
\hline\hline
$N_e$ & $a$ & $n_s$\ \ \ \ \  \ \ \ \ \ \ \ 
& $r$\ \ \ \ \  \ \ \ \ \ \ \ \\
\hline\hline
50$\rightarrow$60 & 0.3 & 0.95965$\rightarrow$0.96641 
& 0.00029$\rightarrow$0.00020 \\
50$\rightarrow$60 & 0.5 & 0.95923$\rightarrow$0.96610 
& 0.00083$\rightarrow$0.00057 \\
50$\rightarrow$60 & 0.7 & 0.95875$\rightarrow$0.96575 
& 0.00168$\rightarrow$0.00115 \\ 
50$\rightarrow$60 & 0.9 & 0.95827$\rightarrow$0.96538 
& 0.00286$\rightarrow$0.00196 \\
50$\rightarrow$60 & 1.0 & 0.95917$\rightarrow$0.96614 
& 0.00328$\rightarrow$0.00224 \\ 
50$\rightarrow$60 & 1.5 & 0.95721$\rightarrow$0.96447 
& 0.00865$\rightarrow$0.00591 \\
50$\rightarrow$60 & 2.0 & 0.95864$\rightarrow$0.96534 
& 0.01472$\rightarrow$0.01013 \\
50$\rightarrow$60 & 3.0 & 0.96292$\rightarrow$0.96780 
& 0.03065$\rightarrow$0.02221 \\
50$\rightarrow$60 & 4.0 & 0.97015$\rightarrow$0.97294 
& 0.04095$\rightarrow$0.03201 \\ 
50$\rightarrow$60 & 5.0 & 0.97708$\rightarrow$0.97852 
& 0.04260$\rightarrow$0.03563 \\
\hline
\end{tabular}
\label{tab1}
\end{table}

\begin{figure}[h]
\centering
\includegraphics[width=0.8\textwidth]{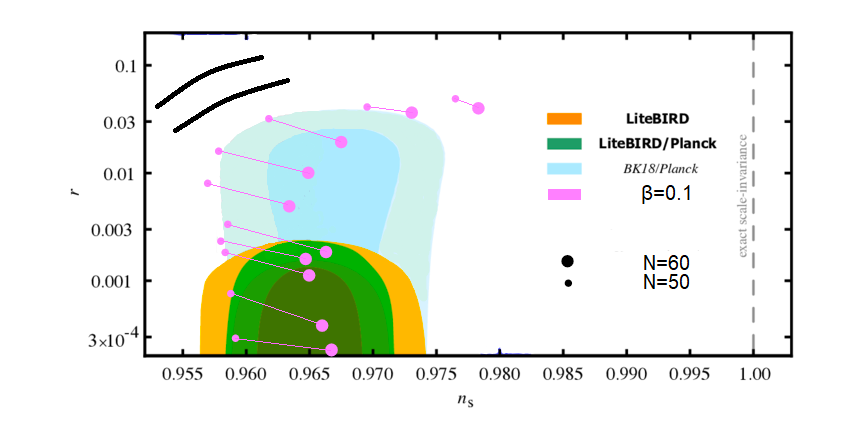}
\caption{The figure represents predictions 
for the tensor-to-scalar ratio 
$r$ and the scalar spectral 
index $n_s$ from the Woods-Saxon model within 
the modified gravity framework. The pink curves show 
the model predictions for $\beta=0.1$ 
with varying $a$, connecting the endpoints for $N_e=50$ 
(small circles) and $N_e=60$ (big circles). 
The shaded regions represent the latest observational 
constraints from the Planck/BICEP/Keck \cite{12}, 
\cite{13}. For comparison, the black curve shows 
the Woods-Saxon potential predictions in the standard 
GR ($\beta=0$), which demonstrates the improvements 
due to the $f(\phi)T$ coupling.}
\label{fig1}
\end{figure}

Figure \ref{fig1} shows $(n_s, r)$ predictions, 
compared with the observational constraints. Predictions 
for $a \lesssim 1$ fall within the  
Planck 2018 $1\sigma$ region for $n_s$ and satisfy 
$r < 0.01$. It exhibits  
good agreement with the current data \cite{12}, 
\cite{13}. Our analysis extends the previous work on 
the modified gravity inflation \cite{32} by incorporating 
the novel Woods-Saxon potential.

\subsection{A comparison with the general relativity}
\label{subsec:GR_comparison}

We study the essential role of the $f(\phi)T$ 
coupling by contrasting results with the 
predictions of the Woods-Saxon 
potential within the standard GR. Setting 
$\beta = 0$ recovers the standard scenario of the 
canonical single-field. In this GR limit, the 
slow-roll parameters simplify to
\begin{align}
\epsilon_V^{\text{GR}} &= \frac{1}{2\kappa} \left( \frac
{V_{,\phi}}{V} \right)^2 = \frac{1}{2\kappa a^2} \left
( \frac{e^{\phi/a}}{1 + e^{\phi/a}} \right)^2, \\
\eta_V^{\text{GR}} &= \frac{1}{\kappa} \frac{V_{,\phi
\phi}}{V} = \frac{1}{\kappa a^2} \frac{e^{\phi/a}(e^
{\phi/a} - 1)}{(1 + e^{\phi/a})^2}.
\end{align}
The number of e-folds is
\begin{equation}
N^{\rm GR} \simeq \kappa\int_{\phi_{\rm end}}^{\phi^{*}}
\!\frac{V}{V_{,\phi}}\,{\rm d}\phi \approx \kappa a^2 
\bigg[ e^{-\phi^{*}/a} - 
e^{-\phi_{\text{end}}/a}- \frac{\phi^{*} - \phi_
{\text{end}}}{a}\bigg].
\end{equation}
The results of the standard GR case are quite 
different from the $f(\phi)T$ case, and they
have been summarized in Table \ref{tabGR}.

\begin{table}[h]
\caption{The predictions of the Woods-Saxon potential 
in the general relativity ($\beta=0$).}
\centering
\begin{tabular}{ccc}
\hline\hline
$a$ & $n_s^{\text{GR}}$ & $r^{\text{GR}}$ \\
\hline\hline
0.3 & 0.9341 $\rightarrow$ 0.9432 & 0.052 
$\rightarrow$ 0.036 \\
0.5 & 0.9365 $\rightarrow$ 0.9450 & 0.056 
$\rightarrow$ 0.038 \\
0.7 & 0.9385 $\rightarrow$ 0.9465 & 0.059 
$\rightarrow$ 0.040 \\
0.9 & 0.9402 $\rightarrow$ 0.9478 & 0.061 
$\rightarrow$ 0.042 \\
1.0 & 0.9410 $\rightarrow$ 0.9484 & 0.062 
$\rightarrow$ 0.042 \\
1.5 & 0.9445 $\rightarrow$ 0.9510 & 0.065 
$\rightarrow$ 0.044 \\
2.0 & 0.9472 $\rightarrow$ 0.9530 & 0.067 
$\rightarrow$ 0.046 \\
\hline
\end{tabular}
\label{tabGR}
\end{table}

The standard predictions of the GR model are 
conclusively ruled out by the modern CMB data. 
The scalar spectral index $n_s^{\text{GR}} 
\approx 0.934-0.953$ entirely lies below the lower 
bound of the Planck 2018 $2\sigma$, which is $0.958$. 
The tensor-to-scalar ratio $r^{\text{GR}} > 0.035$ 
significantly exceeds the preferred upper limit  
$r < 0.01$ (see the combined result of the 
Planck/BICEP-Keck data).

This comparison underscores our central thesis: 
the Woods-Saxon potential in its native form is 
not viable with respect to the current CMB constraints. 
Introducing the non-minimal $f(\phi)T$ 
coupling corrects these deficiencies. The coupling 
alters the inflationary dynamics as shown in 
Eqs. (\ref{teff}), (\ref{sr1}) and (\ref{sr2}), i.e.,  
by flattening the effective potential, 
raising $n_s$ and suppressing $r$. This dynamically 
steers the model predictions into the observationally 
allowed region. 


\subsection{The parameter-space analysis}
\label{subsec:parameter_space}

We extend the analysis beyond the value $\beta=0.1$ 
to examine the robustness of the scenario. We explore full 
observational viability across the ranges of the parameters   
$(a, \beta)$. The detailed numerical scan reveals key 
features which exhibit that the 
model's viability is a robust 
property across a substantial parameter region.

The parameter-space analysis reveals a broad viability 
region spanning $0.4 \lesssim a \lesssim 1.5$ and 
$0.02 \lesssim\beta \lesssim 0.3$. In this region, the 
$(n_s, r)$ predictions fall within the Planck $1\sigma$ 
and $2\sigma$ confidence regions. This extensive viable 
parameter-space demonstrates that successful inflation 
is a generic feature of the model, not a finely-tuned 
special case.

The coupling strength $\beta$ plays a crucial dynamical 
role. For a fixed ``$a$'', increasing $\beta$ systematically 
suppresses the tensor-to-scalar ratio $r$ and increases 
the spectral index $n_s$. This represents the central 
mechanism which rescues the model from exclusion. 
In the standard GR limit ($\beta=0$), the Woods-Saxon 
potential predicts $n_s^{\text{GR}} \approx 0.934-0.953$ 
and $r^{\text{GR}} > 0.035$, in 
which both of them have been 
ruled out by the current data. 
With a nonzero $\beta$, the coupling 
actively corrects these deficiencies, pulling the 
predictions into the observationally favored region 
$n_s\approx 0.959-0.966$ with $r < 0.01$.

The width parameter ``$a$'' controls 
the potential steepness. 
A smaller $a$ (a steeper potential) generally produces 
a higher e-fold number and a lower $r$. However, an
excessively small $a$ can drive $n_s$ too low. Optimal 
parameter-space shows complementary relationship: 
steeper potentials (smaller $a$) require weaker 
coupling (smaller $\beta$), while the broader 
potentials (larger $a$) need stronger coupling 
(larger $\beta$) to achieve the correct spectral index.

Theoretical stability requirements further restrict 
the viable parameter-space. To prevent the unphysical 
ghost instabilities, the perturbation kinetic term 
should remain positive definite, which requires 
$(1 + 2\beta f) > 0$ throughout the inflationary evolution. 
A comprehensive numerical analysis 
excludes the parameter regions with the excessively strong 
coupling. Specifically, $\beta \gtrsim 0.35$ is excluded 
for potentials with $a < 1.0$, and $\beta \gtrsim 0.45$ 
for $a < 1.5$. Our benchmark point $(a=0.7, \beta=0.1)$ 
lies well within the theoretically safe domain,
which confirms both the observational viability 
of the model and its mathematical consistency.

This comprehensive parameter analysis confirms 
that the Woods-Saxon potential, accompanied by 
$f(\phi)T$-gravity, constitutes a robust inflationary 
scenario within the scope of the CMB observables. 
The model is compatible with the observations for a wide and 
natural parameter range. The $f(\phi)T$ coupling 
provides an essential mechanism for achieving 
compatibility with the precision cosmological 
data on the $(n_s, r)$.

\section{Comparison with other modified gravities}
\label{sec7}

Having the viability of the Woods-Saxon model within the 
$f(\phi)T$-framework, we now compare it with other  
modified gravities. This comparative analysis 
highlights the distinctive features and advantages 
of our approach.

\subsection{Relation to the $f(R)$- and 
$f(R,T)$-gravities}

Our $f(\phi)T$-framework represents a specific, 
phenomenologically motivated case of the 
broader $f(R,T)$-gravity class \cite{17}. 
The general $f(R,T)$ action is
\begin{equation}
\label{frt_action}
S_{f(R,T)} = \int {\rm d}^4x \sqrt{-g} 
\left[ \frac{1}{2\kappa}f(R,T) + 
\mathcal{L}_m \right],
\end{equation}
where $f(R,T)$ is a general functional of $R$ and $T$.
Our model corresponds to the choice
\begin{equation}
\label{our_frt}
f(R,T) \rightarrow R + 2\kappa\beta f(\phi)T.
\end{equation}
Thus, for a constant $f(\phi)$, our model reduces to 
an $f(R,T)$-gravity. However, our model 
provides several advantages over the 
general $f(R,T)$-theories:

\begin{itemize}
    \item \textbf{A controlled modification}: 
When $f(\phi) \to 0$, the coupling term $\beta f(\phi)T$ 
naturally vanishes in the post-inflationary era. Hence, it 
automatically satisfies the late-time constraints.
    \item \textbf{A clear physical interpretation}: 
The direct coupling between the inflaton and 
the matter sector during the inflation provides 
a transparent mechanism for altering the 
inflationary trajectory.
    \item \textbf{Predictive power}: 
Specific functional forms for $f(\phi)$ allow precise 
observational predictions and admit a direct comparison
with the data.
\end{itemize}

Comparing to the $f(R)$-gravity \cite{15}, 
our approach retains the Einstein-Hilbert term,
while introduces a matter couplings   
provides a complementary modification mechanism.

Whereas $f(\phi)R$-theories (of the Brans-Dicke type) 
modify the gravitational sector, our $f(\phi)T$ 
approach directly couples the inflaton to matter:
\begin{equation}
\label{comparison_eq_detailed}
\frac{\mathcal{L}_{f(\phi)R}}{\sqrt{-g}} 
= f(\phi)R \quad \text{vs.} 
\quad \frac{\mathcal{L}_{f(\phi)T}}{\sqrt{-g}} 
= \beta f(\phi)T.
\end{equation}
This provides a complementary mechanism for altering the 
inflationary dynamics which operates without modifying the 
gravitational constant or introducing additional 
scalar degrees of freedom.

Table \ref{tab2} provides a comparative overview 
of the inflationary scenarios within 
the standard and modified gravity frameworks.

\begin{table}[h]
\centering
\caption{Comparison of the inflationary predictions 
across different modified gravities}
\label{tab:comparison}
\begin{tabular}{p{3.5cm}ccc}
\hline\hline
\textbf{Model} & \textbf{Typical $n_s$} 
& \textbf{Typical $r$} & 
\textbf{Distinctive Features} \\
\hline\hline
Standard GR+Woods-Saxon & 0.94-0.95 & 0.02-0.04 
& Excluded by the Planck \\
\hline
$f(\phi)T$+Woods-Saxon (this work) & 0.959-0.966 
& $0.01\sim 0.001$ & \textbf{Planck-compatible}, 
minimal modification \\
\hline
Starobinsky $f(R)$ \cite{15} & 0.965 & 0.003 
& No direct matter coupling \\
\hline
Higgs Inflation \cite{33}, \cite{34} & 0.967 & 0.003 
& Non-minimal coupling to $R$ \\
\hline
$\alpha$-attractors \cite{35}-\cite{39} & 0.960-0.967 
& $<$0.07 & Universal predictions \\
\hline
$f(\phi)R$+Woods-Saxon & 0.945-0.955 & 0.02-0.03 
& Still disfavored \\
\hline
\end{tabular}
\label{tab2}
\end{table}

Within the ordinary general relativity, 
the Woods-Saxon potential predicts a relatively 
low scalar spectral index and a large 
tensor-to-scalar ratio. These predictions 
place it outside the Planck 2018 confidence region. 
In contrast, the same potential, embedded in 
$f(\phi)T$ framework, yields 
$n_s\simeq0.959\text{--}0.966$ and $r<0.01$, which shows  
a good agreement with the Planck/BICEP-Keck bounds on 
these observables.

\subsection{Reheating in the $f(\phi)T$ framework}

We now briefly address the reheating epoch within the 
$f(\phi)T$ framework. 
Reheating in this class of theories has been studied in 
the previous works, such as Ref. \cite{40}.
They have analyzed 
the post-inflationary dynamics and the efficiency of 
energy transfer from the inflaton to the Standard Model 
degrees of freedom.

For the Woods-Saxon potential which we considered, the 
inflaton rolls from the plateau region at $\phi \to -\infty$ 
toward $\phi \to +\infty$, where the potential vanishes. 
The shape of the potential near its minimum can significantly 
affect the reheating temperature and the efficiency of 
particle production. However, the 
specific detail of the reheating 
depends on the inflaton's couplings to other fields, which 
are not specified in our minimal model. Therefore, we 
expect that the general conclusions 
of Ref. \cite{40}, regarding 
the viability of the reheating in 
$f(\phi)T$ theories, will apply 
qualitatively to our setup. In fact, a detailed analysis for 
this specific potential is beyond the scope of the present 
work. We leave a comprehensive study of the reheating in the 
Woods-Saxon $f(\phi)T$ model for the future investigation.

\subsection{Other advantages}

Our framework combines several desirable features 
which distinguishes it from the other modified gravity 
inflation:

\begin{itemize}
    \item \textbf{Novel potential}: The potential of 
the model is the first application 
of the Woods-Saxon potential in the cosmology.
    \item \textbf{Minimal modification}: The added single  
term in the action maintains theoretical simplicity, while 
we achieve the observational compatibility.
    \item \textbf{Graceful exit}: Automatic returning 
to the standard GR through the asymptotic behavior 
$f(\phi) \to 0$ as $\phi \to 0$.
    \item \textbf{Comprehensive analysis}: 
Detailed treatment of the CMB predictions and the  
theoretical consistency.
    \item \textbf{Testable predictions}: Specific 
relationships between the model parameters and the  
observables constrainable by the future experiments.
\end{itemize}

\section{Conclusions and discussion}
\label{sec8}

We constructed a viable and compelling inflationary 
scenario by synthesizing the empirically-motivated Woods-
Saxon potential, accompanied by the modified gravity
$f(\phi)T$. This cross-disciplinary approach 
yields a model that is both theoretically consistent 
and observationally precise with respect to the CMB two-
point statistics.

One key result is that the non-minimal coupling 
in the $f(\phi)T$-gravity is not a minor 
adjustment but an essential ingredient for the 
observational viability of the Woods-Saxon potential with 
respect to the $(n_s, r)$ constraints. 
In fact, the potential in its native and general 
relativistic form is conclusively ruled out by the CMB data. 
Thus, the introduction of the coupling governed by a 
simple functional $f(\phi)$, that 
vanishes in the post-inflation, flattens the effective 
potential. This mechanism successfully pulls the predictions 
for the spectral index and tensor-to-scalar ratio 
into the favored region by the Planck and BICEP/Keck 
data, i.e. $n_s \approx 0.959\text{--}0.966$ 
and $r < 0.01$, for a broad region of the parameter-space 
($a \lesssim 1$, $\beta \sim 0.1$).

This work underscores several key insights. 
First, it demonstrates that the phenomenological 
potentials from other areas of 
physics can serve fertile ground for the 
cosmological models-building, they are 
paired with an appropriate 
gravitational framework. Second, it highlights 
the $f(\phi)T$-framework as a particularly 
attractive and minimal modification to the GR.
For example, it provides a clear mechanism for 
altering the inflationary dynamics. Besides, it
guarantees a graceful exit into the standard cosmology. The 
model's predictions are highly testable. Precisely, 
the specific relationships, that we derived 
between its parameters and 
the observables, will be sharply constrained by the 
next-generation of the CMB experiments.

Regarding the post-inflationary dynamics, we  
briefly discussed the reheating within the 
$f(\phi)T$ framework. For the Woods-Saxon potential, 
the inflaton rolls from the plateau region toward the 
minimum where the potential vanishes. The 
general conclusions of Ref.~\cite{40}, regarding the 
viability of the reheating in the $f(\phi)T$ theories, are 
expected to apply qualitatively. However, a detailed analysis 
via this specific potential remains an open question 
for future work.

Finally, combining the Woods-Saxon 
potential with the $f(\phi)T$-gravity opens a new 
pathway for the inflationary model-building.  
It also provides a tangible example that the 
insights from the nuclear physics can illuminate our 
understanding of the early universe.

\textit{A note on the scope of our analysis.} 
We emphasize that our conclusions regarding 
the viability of the model are based on the 
inflationary dynamics and the two-point CMB 
observables $(n_s, r)$. A comprehensive assessment 
of the model requires some additional 
aspects, including the reheating (briefly discussed in 
Section 7), the normalization of the scalar power 
spectrum, the running of the spectral index, non-
Gaussianities, and a complete stability analysis. 
These topics are beyond the scope of the present 
work and are left for the future investigations. We 
regard this work as a first step in establishing the 
Woods-Saxon potential within the $f(\phi)T$ gravity, 
and we hope that future studies will explore these 
important directions.

\newpage

\end{document}